**Autonomous scanning probe microscopy with hypothesis learning: Exploring the physics of domain switching in ferroelectric materials**

Yongtao Liu,[1a] Anna Morozovska,[2] Eugene Eliseev,[2,3] Kyle P. Kelley,[1] Rama Vasudevan,[1] Maxim Ziatdinov,[1,4,b] and Sergei V. Kalinin[1,c]

[1] Center for Nanophase Materials Sciences, Oak Ridge National Laboratory, Oak Ridge, TN 37922, United States

[2] Institute of Physics, National Academy of Sciences of Ukraine, 46, pr. Nauky, 03028 Kyiv, Ukraine

[3] Institute for Problems of Materials Science, National Academy of Sciences of Ukraine, Krjijanovskogo 3, 03142 Kyiv, Ukraine

[4] Computational Sciences and Engineering Division, Oak Ridge National Laboratory, Oak Ridge, TN 37831, United States

We report the development and implementation of a hypothesis-learning-based automated experiment, in which the microscope operating in the autonomous mode identifies the physical laws behind the material's response. Specifically, we explore the bias-induced transformations that underpin the functionality of broad classes of devices and functional materials from batteries and memristors to ferroelectrics and antiferroelectrics. Optimization and design of these materials require probing the mechanisms of these transformations on the nanometer scale as a function of the broad range of control parameters such as applied potential and time, often leading to experimentally intractable scenarios. At the same time, often the behaviors of these systems are understood within potentially competing theoretical models, or hypotheses. Here, we develop a hypothesis list that covers the possible limiting scenarios for the domain growth, including thermodynamic, domain wall pinning, and screening limited. We further develop and experimentally implement the hypothesis-driven automated experiment in Piezoresponse Force Microscopy, autonomously identifying the mechanisms of the bias-induced domain switching. This approach can be applied for a broad range of physical and chemical experiments with relatively low-dimensional control parameter space and for which the possible competing models of the system behavior that ideally cover the full range of physical eventualities are known or can be created. These include other scanning probe microscopy modalities such as force-distance curve measurements and nanoindentation, as well as materials synthesis and optimization.

[a] liuy@ornl.gov
[b] ziatdinovma@ornl.gov
[c] sergei2@ornl.gov

Bias-induced transformations underpin the functionality of broad classes of functional materials and devices. Electrochemical reactions and intercalation underpin the functionalities of multiple classes of energy storage materials[1, 2] and memristive electronics.[3-6] Polarization switching in ferroelectrics is broadly used in non-volatile random access memories,[7-9] ferroelectric tunneling barriers,[10, 11] multiferroic,[12, 13] and field-effect devices.[14] Similarly, domain wall dynamics strongly couple to the electromechanical and electrooptical responses in these systems.[15, 16] Bias-induced phase transitions in antiferroelectrics underpin the applications in energy storage devices and tunable electronics.[17]

In virtually all cases, the bias-induced transformations, including polarization switching or electrochemical processes, are controlled by the interplay of the nucleation and phase transformation front propagation and pinning. These phenomena, in turn, are strongly linked to spatial inhomogeneities and defects that can act as nucleation centers for new phases, pinning centers for moving transformation fronts, and affect the reaction pathways, e.g., phase or crystallographic orientation selection. Correspondingly, much attention has been focused on exploring these phenomena locally, via electron microscopy, focused X-Ray, and scanning probe microscopy (SPM), as well as combined EM-SPM modalities.[18, 19] In particular, scanning probe microscopy based methods offer an experimental framework where bias-induced transformation can be probed over a dense rectangular grid of spatial locations, and machine-learning or physics-based analysis of the resultant dataset allows visualizing the salient features of materials behavior in space.[20, 21]

The mechanistic insights in these phenomena are traditionally derived by probing changes in material structure and functionality as a function of control parameters such as applied potential and time. For macroscopic global measurements, such studies - as implemented in classical electrochemical and ferroelectric device characterizations techniques such as PUND and first-order reversal curve measurements for ferroelectrics or potential intermittent titration (PITT)[22] for electrochemical system - often result in extremely lengthy measurements. For local characterization, with few exceptions[23, 24] this leads to the experimentally intractable scenarios due to the need for probing very high dimensional parameter spaces. These considerations underscore the need to develop automated experiment workflows that allow exploring relevant behaviors in a targeted manner, avoiding direct grid sampling of parameter spaces.

The need for automated experimentation has been recently recognized across multiple areas of instrument-based sciences, including X-Ray scattering, electron, and scanning probe microscopy.[25-27] Similarly, rapid growth in automated synthesis platforms, including computer-controlled synthesis,[28] fully automated labs,[29] microfluidic systems,[30] and combined human-high throughput experimentation workflows[31] necessitates development of algorithms for navigating multidimensional compositional or processing spaces. Notably, the (initial) requirements for automated experiment in microscopy and synthesis are close, and allow for the use of classes of algorithms based on Bayesian Optimization (BO).[32-34]

The important limitation of the classical BO strategies with the Gaussian process is the use of the non-parametric kernel-based models. In this case, the internal correlations across the data space are used to select the locations for a new experiment. However, these models do not contain any specific physical assumptions or relationships. Hence in many cases, the efficiency of BO based active learning methods is only within an order of magnitude from classical grid search-based strategies. At the same time, often the behaviors of these systems are understood within potentially competing theoretical models, or hypotheses.

Recently, we have introduced the approach for the physics-informed BO in automated experiments, referred to as hypothesis learning.[35] In this approach, a list of possible models (hypotheses) of the system behavior is established prior to the automated experiment. The hypotheses in this case are the analytical expressions (or other fast computational schemes), with the partial knowledge of the associated parameters in the form of Bayesian priors formed based on the analysis of physics of observed phenomena. During the experiment, the algorithm aims to narrow down the range of possible hypotheses following a certain optimization policy, i.e., it tries to establish the best model of the system's behavior within the smallest number of steps. The thus identified model will represent the mechanism of the observed physical phenomena. Ideally, the list of hypotheses will enumerate possible scenarios for materials behavior; however, if the correct model is not a part of the list, the algorithm reverts to the structured model closest to the ground truth behavior or adopts a structureless Gaussian prior.

Here, we illustrate the hypothesis learning-based automated experiment for the explorations of the domain switching mechanisms in classical ferroelectric materials. While shown for a model system, this approach is more general and can be broadly used for exploration of other

SPM based electrochemical reactions, automated experimentation in scattering, microscopy, and materials synthesis.

## I. Principles of hypothesis learning

The bedrock element of physical sciences is the validated set of quantitative symbolic relationships between the physical parameters, numerical constants, and observables. Examples range from the fundamental laws of the Newtonian mechanics to expressions defining current-voltage relationships in semiconductors. In certain cases, these relationships are derived from fundamental laws and symmetries. In others, they represent a useful empirical generalization, valid under specific conditions. Using these relationships underpins all areas of physics research, and deriving these relationships is often equated with the understanding of the relevant physical mechanisms.

The research process often involves iterative cycles between the acquisition of the experimental data and its interpretation in terms of specific models. In some cases, the underlying symbolic relationships are discovered via exploratory data analysis, with the subsequent interpretation based on the functional form of derived relationships. In other cases, a number of competing models can be derived based on prior knowledge and fundamental physical laws, and the model best matching the experiment is selected to represent the relevant physics. In all cases, models include not only the symbolic form *per se*, but also the expected values of the internal parameters that are known with different degree of certainty, naturally cast in the Bayesian inference (BI) framework.

Hypothesis learning is developed as an approach to implement this iterative cycle as a part of the automated experiment.[35] Here, the automated experiment generally refers to sequential (or batch) measurements of a target functionality over pre-defined parameter space. The scalarizer function reduces the (potentially vector-valued) functionality to a single scalar, defined to represent a measure of the experimentalist's interest in a specific physical property or response. Several possible hypotheses describing system's behavior are available to complement an automated experiment. The hypothesis generally refers to a model predicting the functionality of interest (or its scalarized form) over the parameter space. Ideally, the list of models reflects the full list of possible mechanisms active within the material, and fully covers the possible physical scenarios. The key requirement of the model is the ease of calculation, as necessary to perform

Bayesian evaluations based on Markov Chain Monte Carlo techniques. Here, for convenience, we use the hypotheses in the symbolic equation form; however, this requirement can be relaxed to numerical models. The symbolic expression and associated prior distributions of parameter define a single hypothesis. A list of possible hypotheses with associated prior probabilities (which may or may not include the ground truth one) is a second component of hypothesis learning.

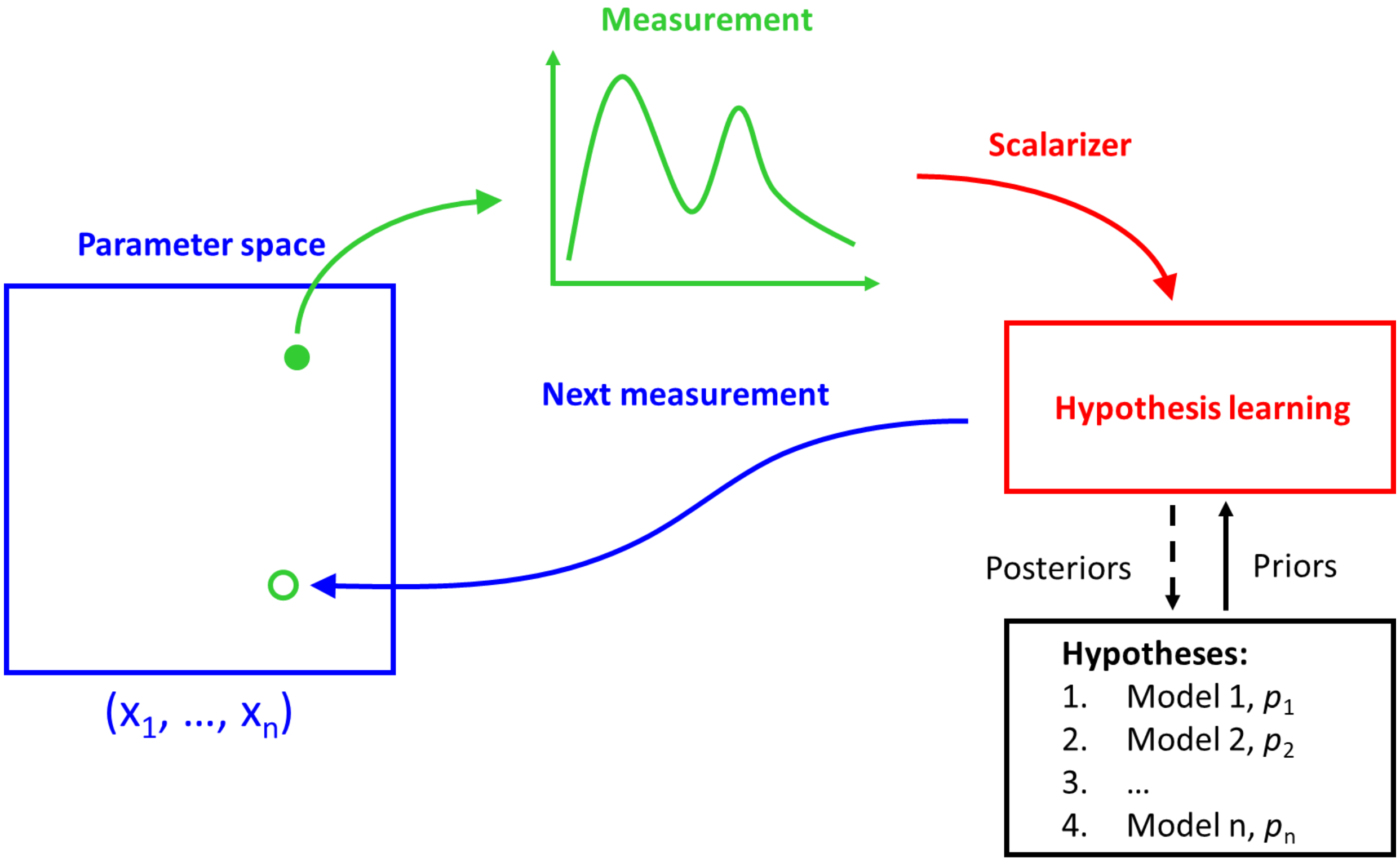

**Figure 1.** Schematics of the hypothesis learning in automated experiment. The measurement is performed in a selected location(s) in the parameter space. These can be control parameters of experiment, concentrations in the phase diagram, or image plane. The measurement result is converted to a scalar measure of interest. The hypothesis learning agent uses the measurement to establish the posterior probabilities of a sampled model and select next location(s) in parameter space for measurements. Note that models corresponding to different physical scenarios (hypotheses) return the scalar values derived via the same scalarizer as the experiment.

During the hypothesis learning, the agent performs the experiment, returning the scalarized value of the functionality of interest in a selected point of the parameter space. The measured values are used to perform the BI on the list of probabilistic models (hypotheses) wrapped into the structured Gaussian processes,[36] generating posterior probabilities of the models' parameters. The latter is used to obtain posterior predictive uncertainties over the unmeasured points of the parameters space. The model that produced the lowest predictive uncertainty is assigned a positive reward value and is used to sample the next measurement point according to a pre-defined acquisition function. Because running BI for every model on the list at each step is computationally expensive, we do it only for several steps ("warm-up phase") and then switch to the epsilon-greedy policy for sampling a single model (hypothesis) at each step. If the sampled model reduces/increases the predictive uncertainty compared to the previous step, it receives a positive/negative reward.

## II. Mechanisms for ferroelectric domain growth

The phenomenological mechanisms for ferroelectric domain growth in PFM have been extensively explored for over two decades.[37-39] The early studies have established the phenomenological relationship between the size of the formed domain and the parameters of the bias pulse applied to the tip.[40-43] The initial theoretical analyses of the domain switching were based on either purely thermodynamic considerations in the rigid ferroelectric[44-46] or Ginzburg-Landau approximations,[47, 48] or analysis of the domain wall motion in the electrostatic field of the probe.[43, 49, 50] At the same time, these analyses have demonstrated that thermodynamics of polarization switching strongly depends on the effectiveness of the screening process on the top surface. Correspondingly, the kinetics of the process can be limited by the screening process, rather than intrinsic material behavior. From the experimental perspective, ample evidence exists towards the role of screening charge dynamics in switching through observations of charge injection,[51, 52] back switching and formation of bubble domains,[53-57] chaotic switching dynamics[58] and formation of complex domains,[59] vortices and skyrmions.[60-62]

Hence, despite its apparent simplicity, domain switching in ferroelectrics is a complex process that is affected by the intrinsic thermodynamics of the domain formation, domains wall pinning in the spatially non-uniform probe field, and screening charge generation and dynamics. Any of these can serve as a process limiting stage, forming ideal setting for hypothesis learning applications. Importantly, that at the mesoscopic level these mechanisms provide the full range of possible physical eventualities, and thus the hypothesis list is assumed to be complete.

Here, we enumerate the hypotheses for the domain growth based on: (a) thermodynamic control in the presence or absence of surface screening charges, (b) kinetic control of domain wall motion via pinning, or (c) kinetic control via screening charge dynamics. This list of possible limiting factor is exhaustive for the domain formation in PFM, and hence the selected hypotheses are expected to cover the full list of experimental eventualities. The analysis of the laws of domain growth in these cases was carried out by a number of groups over the last two decades,[63-68] and below we give only corresponding approximate expressions. The detailed voltage dependences of the equilibrium domain sizes (a)-(c) are described in details in **Appendix A. Figure 2** illustrate schematically the domain nucleation models.

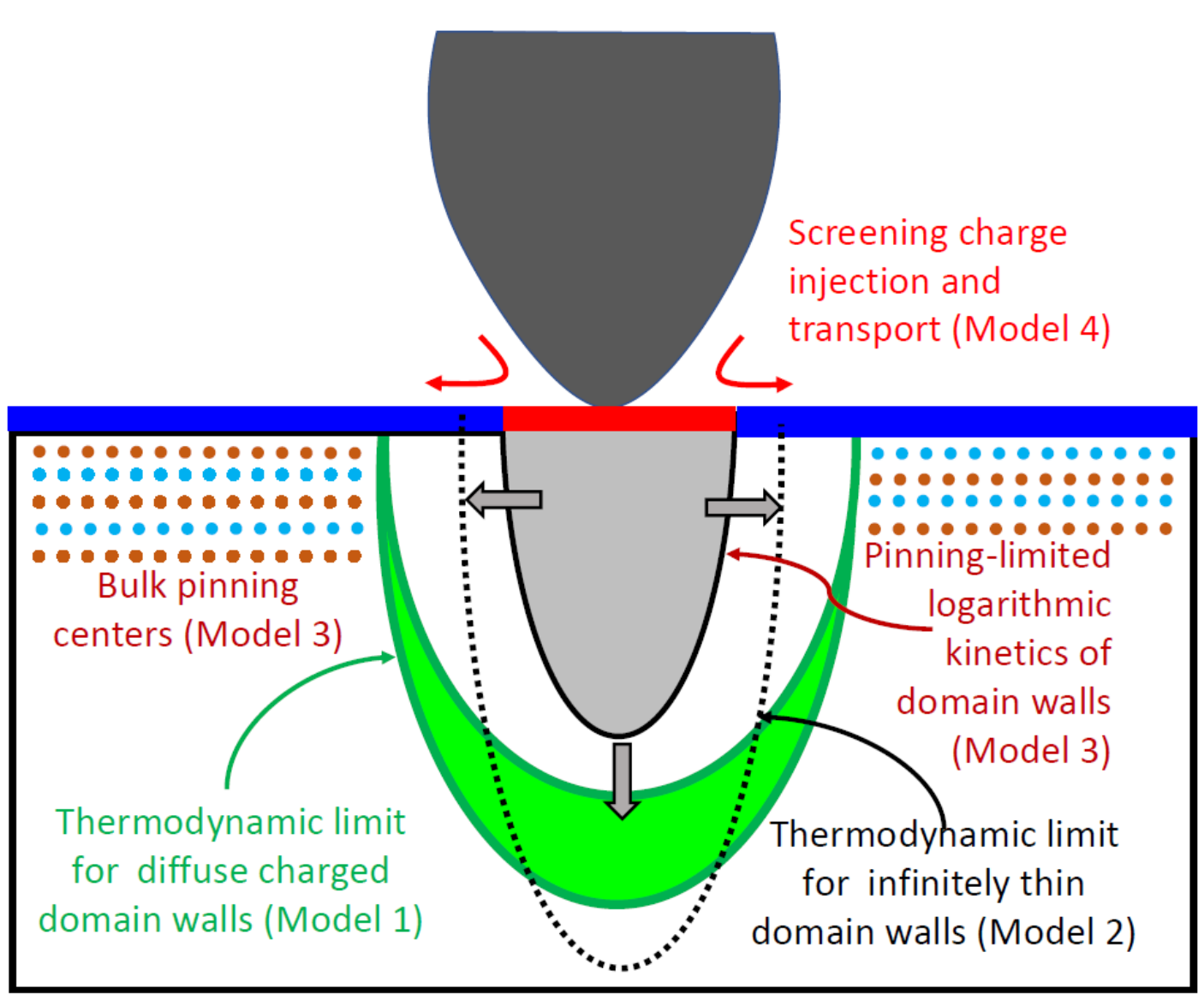

**Figure 2.** Schematics of the domain nucleation models. Model 1 shows thermodynamic limit for diffuse domain walls, which are thicker when becomes charged. Model 2 shows thermodynamic limit for infinitely-thin domain walls. Model 3 is limited by domain wall pining in the bulk, giving rise for the logarithmic kinetics. Model 4 is limited by the injection and transport of surface charged species that are necessary for the polarization screening.

Here, **Model I** corresponds to the partial internal screening of depolarization field by space charge carriers and domain wall thickening. For the case the equilibrium domain radius depends on the tip voltage as:

$$r(V) \approx r_{cr} + d\sqrt{\left(\frac{V}{V_c}\right)^{2/3} - 1}, \tag{1}$$

meaning jump at $V_c$ corresponding to the first order phase transition, and further growth as roughly as $(V/V_c)^{1/3}$. For complete screening, the jump disappears, and the equilibrium domain radius depends on the tip voltage as $r(V) \approx d\sqrt{(V/V_c)^{2/3} - 1}$, i.e., as the second order phase transition scenario. Note that in the Bayesian setting these models are similar and can be defined by tuning the prior distributions on parameter $r_{cr}$.

**Model II.** Landauer-Molotskii (LM) approach of an infinitely thin domain walls, very prolate domains and their breakdown (see **Fig. 1c**).[69, 70] For the case the equilibrium domain radius depends on the tip voltage as:

$$r(V) = r_{cr} + r_0\sqrt[3]{\left(\frac{V}{V_c}\right)^2 - 1}. \quad (3)$$

Again jump $V_c$ corresponding to the first order phase transition, and further growth as $\left(\frac{V}{V_c}\right)^{2/3}$.

**Model III.** The alternative to the thermodynamic control of the domain size is the kinetic control, in which case the domain size is determined by the kinetics of the domain wall motion. From general theory of the disordered media, that the domain wall velocity in the uniform field follows the classical dependence including pinning, creep, and then depinning and linear motion.[41, 71-73] The available kinetic models[50, 74] for the domain wall velocity $v(r)$ in a ferroelectric with pinning relate an acting electric field $E$, threshold field $E_{th}$ and as $v(r) \approx v_0\, exp[-(E_{th}/E)^{\mu}]$, where $\mu$ is a positive exponential factor, which is typically close to unity. Using the simplest form for a normal component of the tip field, $E_z(r,0) = \frac{Vd^2}{\gamma(r^2+d^2)^{3/2}}$, where $\gamma$ is a dielectric anisotropy factor, $d$ is the effective tip size, $r$ is a surface distance from the tip axis, and $V$ is the bias applied between the tip and the bottom electrode, the approximate solution for the time dependence of the domain radius can be derived as (Appendix A):

$$r(t) \approx \left(\frac{V}{\beta}\right)^{1/3} ln\left[1 + \left(\frac{\beta}{V}\right)^{1/3} v_0 t\right], \quad (4)$$

Where the parameter $\beta = \gamma E_{th}/d^2$. Expression (4) describes a slow logarithmic creep of the domain wall, at that $r(V) \sim V^{1/3}$ at high voltages. For $(\beta/V)^{1/3} v_0 t \gg 1$, we obtain that $r(t) \sim (V/\beta)^{1/3} ln[v_0 t]$. However, the lateral growth stops at equilibrium domain sizes after the pulse ending, which can be calculated from thermodynamic description.

Finally, as a **Model IV**, we consider the case where the domain growth is limited by the transport of the screening charges across the sample surface. Here, we note that, in general, polarization switching requires almost complete compensation of the polarization charges by screening charges. If screening charges are abundant, the domain is determined by switching thermodynamics (Model 1, 2) or wall pinning (Model 3). If the screening charges are slow and sparse, the domain growth is limited by the charge injection. The experimental evidence towards this behavior was obtained by Yudin *et al*,[43] and also indirectly via observations of phenomena such as chaotic domain switching.[58, 59]

The simple consideration of the mass and charge balance suggest that in the PFM experiment the screening charges can be generated only at the tip-surface junction. In this case, assuming general power law voltage dependence of the generation rate and diffusional or drift transport of charge species, the kinetics of the domain wall growth can be described as

$$r(t) \approx V^{\alpha}\tau^{\beta}, \tag{4}$$

where, in the depletion approximation, $\alpha$ is close to 1 and $\beta$ is ½.

## III. Experimental realization of hypothesis learning

As a model ferroelectric system, we use a fully relaxed 80 nm thick $BaTiO_3$ (BTO) thin film (see Methods). A representative PFM image and domain writing is shown in Figure 3. Figure 3a is the surface topography showing uniform geometry with periodical terrace structures. Figure 3b-3c are out-of-plane band-excitation PFM (BEPFM) amplitude and phase images of the original sample, respectively, and Figure 3d shows the corresponding resonance frequency related to local elastic property. The BTO film shows a down-polarized pristine state (Figure 3c). The polarization can be switched by applying a DC bias via AFM tip. Shown in Figure 3e-g are out of plane BEPFM amplitude, phase, and frequency images showing pre-poled areas by applying 5 V DC bias via AFM tip.

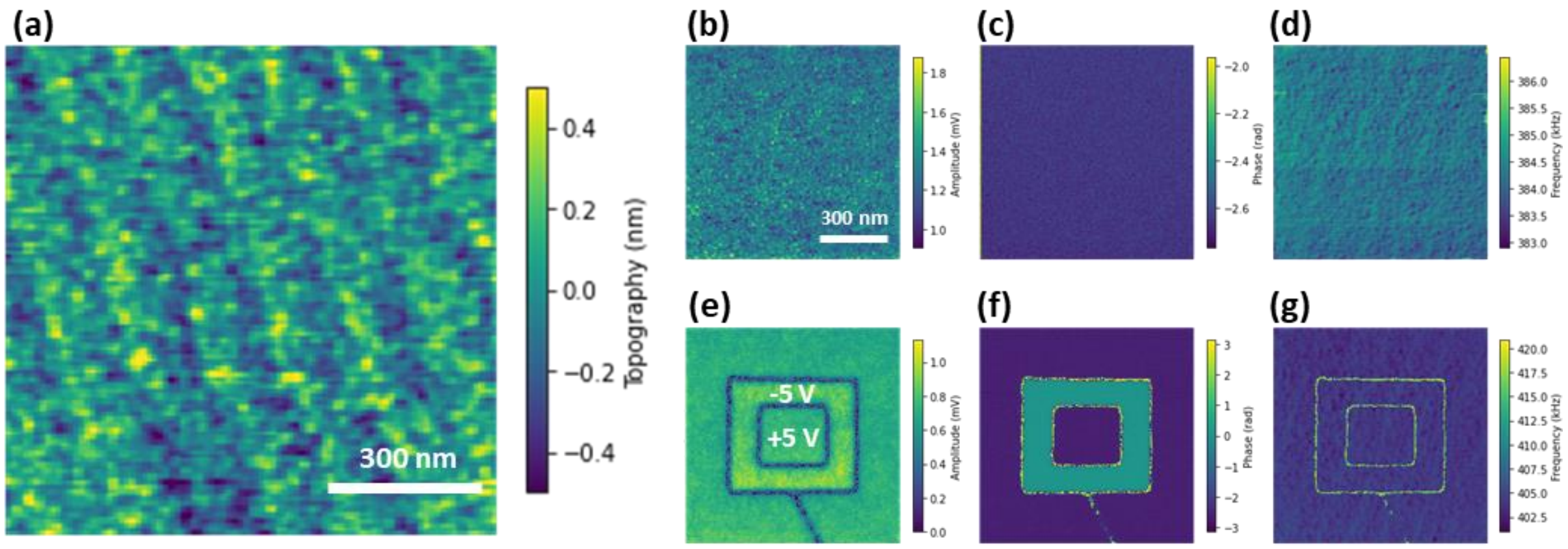

**Figure 3.** BEPFM of BTO sample. **(a)** Topography with periodical terrace structure. **(b-d)** BEPFM amplitude, phase, and frequency images of the pristine state. **(e-g),** BEPFM amplitude, phase, and frequency images of pre-poled areas by 5 V DC bias.

To realize hypothesis-learning-based automated experiment, we developed and deployed a workflow shown in Figure 4a integrating multiple software (including LabView, Jupyter Notebook, Google Colaboratory, and Igor) and hardware (including a National Instruments DAQ card, a Field Programmable Gate Arrays (FPGA), and Asylum Research Cypher microscope). To perform a BEPFM measurement, the measurement location (equivalent to tip location) is controlled by FPGA and BEPFM data is acquired by National Instruments DAQ card. To apply a pulse bias for writing domain, FPGA move the tip to the target location (center of the experiment area) and apply the pulse bias to tip.

The workflow of hypothesis-learning automated BEPFM is shown in Figure 4b. The experiment is started with initializing the measurement area by applying a DC bias to uniformly pole this area toward the same direction. Then, domain writing and imaging are performed in this area. The acquired BEPFM data is analyzed by a threshold filter to detect the written domain, and the domain size is determined by the minimum closed circle in which the written domain lies either inside the circle or on its boundaries. Next, this domain size and writing parameters, as well as all previous domain sizes and writing parameters, is fed to the hypothesis learning model to predict the next writing parameters. Simultaneously, when the hypothesis learning algorithm is performing training and prediction, the workflow also controls the microscope to back-switch the measurement area (erase the domain) in order to be ready for next writing iteration. Then, the predicted writing parameters will be fed to the measurement workflow and the writing/imaging process will be performed again. for this workflow, we also added a function to check the BEPFM data quality, such that a poor dataset will be discarded, and the same measurement will be repeated. This checking process ensures that all BEPFM data and, consequently, domain sizes are comparable.

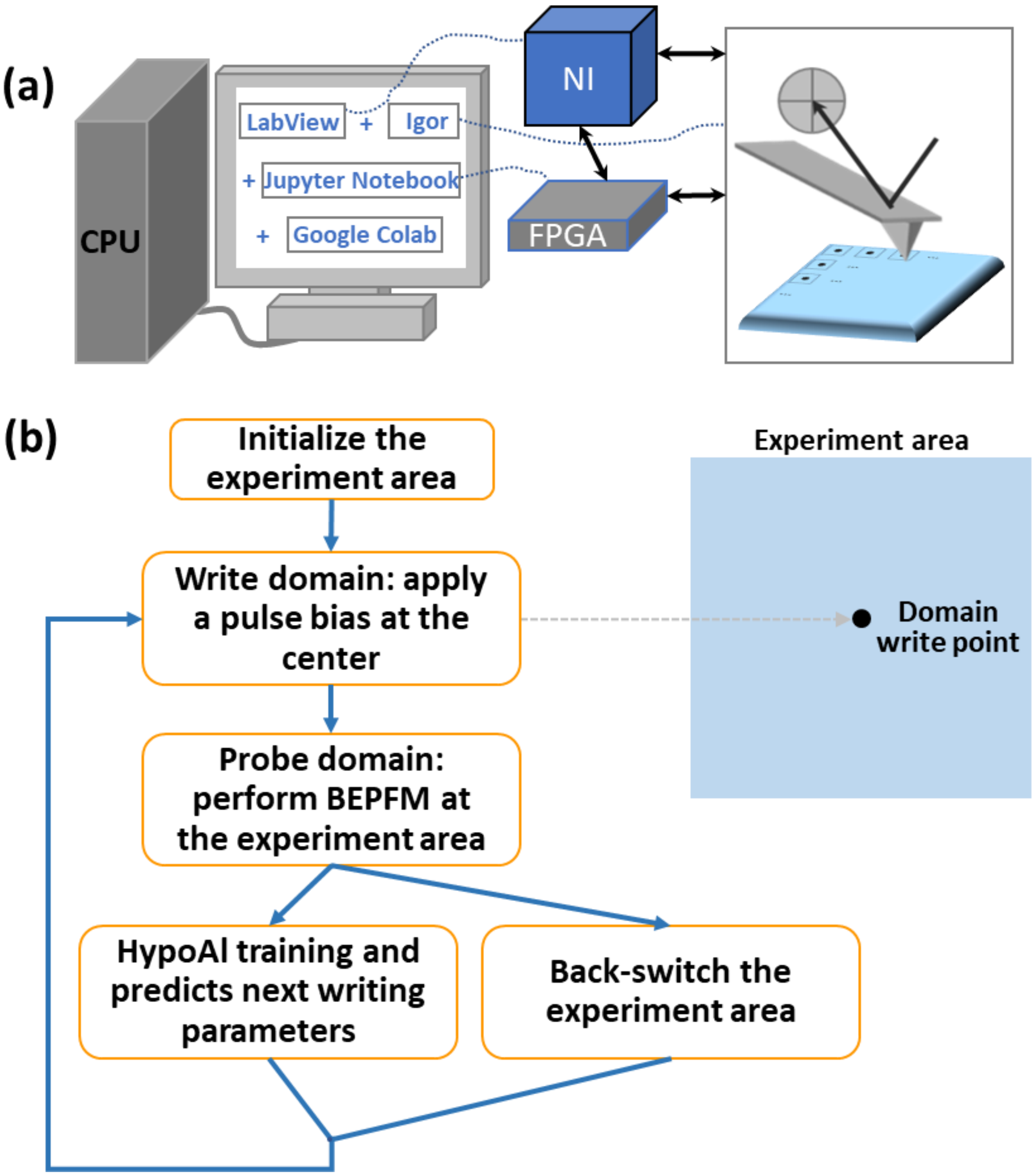

**Figure 4.** Hypothesis learning automated BEPFM system and workflow. **(a)** a schematic showing the integrated system for hypothesis learning automated BEPFM; **(b)** hypothesis-learning-based automated BEPFM workflow.

For the hypotheses list, we have chosen four models following our discussion in Section II. Equations corresponding to these models are shown in Table 1.

**Table 1.** Model equations and priors used in the hypotheses-driven automated experiment. More details about these models are available in Supplementary Materials Appendix. A.

| | **Model Equation** | **Model Priors** |
|---|---|---|
| **Model I** | $r(V) = r_{cr} + r_0\sqrt{\left(\frac{V}{V_c}\right)^{2/3} - 1}$ | $r_{cr} \sim Normal(0,1)$<br>$r_0 \sim LogNormal(0,1)$<br>$V_c \sim LogNormal(0,1)$ |
| **Model II** | $r(V) = r_{cr} + r_0\sqrt[3]{\left(\frac{V}{V_c}\right)^{2} - 1}$ | $r_{cr} \sim Normal(0,1)$<br>$r_0 \sim LogNormal(0,1)$<br>$V_c \sim LogNormal(0,1)$ |
| **Model III** | $r(V,t) = V^{\alpha} \log \tau$ | $\alpha \sim Uniform(0.33, 1.2)$ |
| **Model IV** | $r(V,t) = V^{\alpha}\tau^{\beta}$ | $\alpha \sim Uniform(0.8, 1.2)$<br>$\beta \sim Uniform(0.33, 1.2)$ |

Shown in Figure 5 are the hypotheses-learning-based automated BEPFM results. In this experiment, 18 random writing parameters (5% of the writing parameters library) were selected to perform the initial domain writing experiment to provide initial ("seed") points for hypothesis learning. The obtained domain sizes along with corresponding writing parameters were used as initial training data for the hypothesis learning algorithm. Then, 40 measurements were performed using writing parameters predicted by the algorithm. Figure 5a shows a few examples of the domains written in the BTO thin film, along with the binary domain images and domain size automatically detected by the workflow. It indicates that in this experiment both the writing bias and writing time affect the domain size. In Figure 5b, all obtained results are shown as domain sizes as a function of write bias and time. Clearly, the larger bias and longer time result in increased domain sizes. Figure 5c shows the usage times of each model in the 40-steps hypotheses learning, where the most often sampled model is Model III. Figure 5d shows the evolution of rewards of each model. Model III 'won' the most rewards and its reward values steadily grow in the latter part of the experiment (this explains why it was selected more frequently than other models).

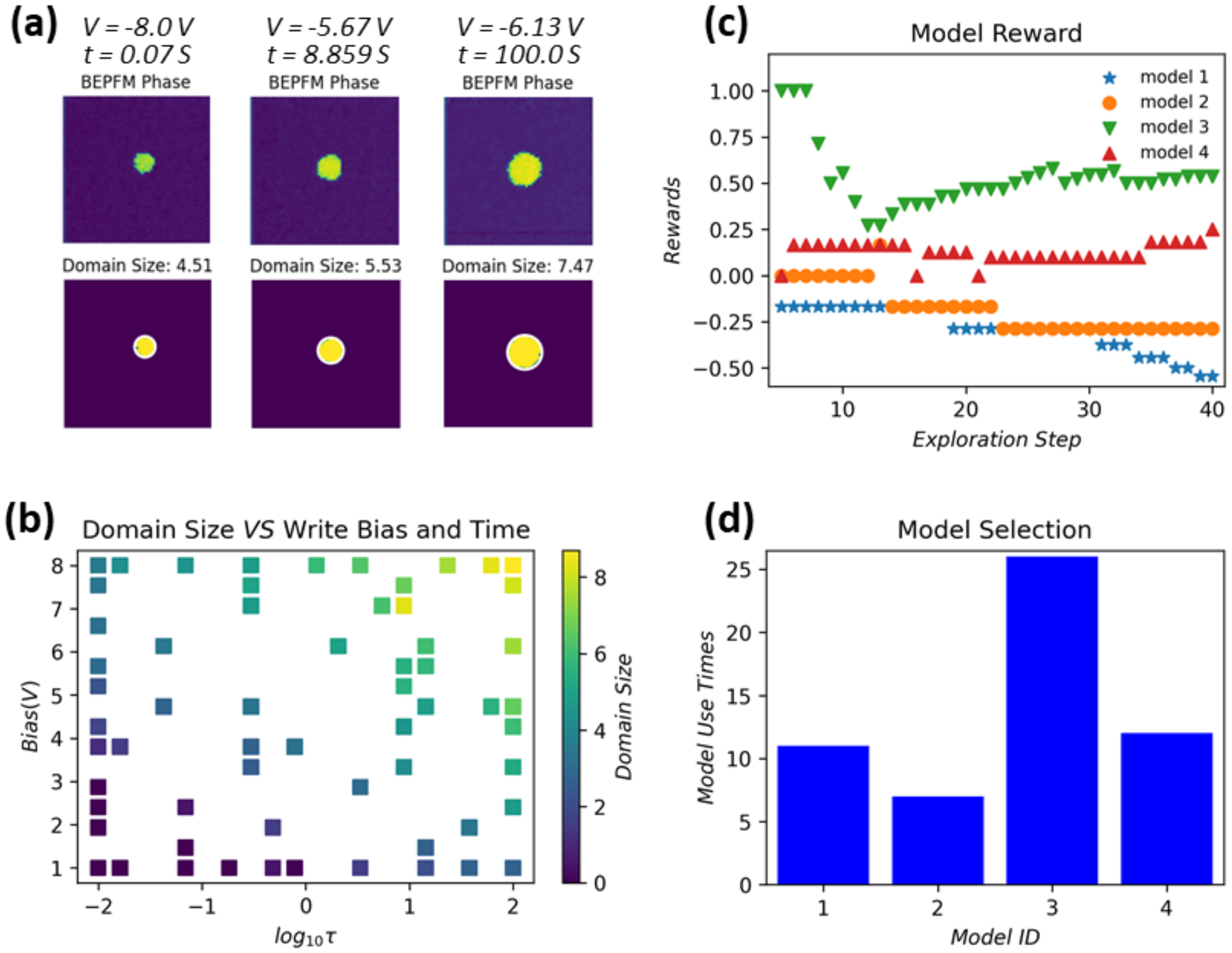

**Figure 5.** Hypotheses-learning-based automated BEPFM experiment results. **(a),** three examples of domains written by using different bias and time. The top row are BEPFM images showing the domains and the bottom row are corresponding binary images with domain size detected automatically by the automated workflow. Note that all BEPFM results and domain binary images are shown in Supplementary Video as a function of the measurement step. **(b),** Domain size as a function of writing parameters. **(c),** Model selection in hypotheses-learning, in which model III was selected more often than other models. **(d),** Model rewards during the hypotheses learning after the initial 5-step warm-up phase during which all models were evaluated at each step. Model III gained a much larger reward than other models, and its reward gradually increased at the latter part of the experiment.

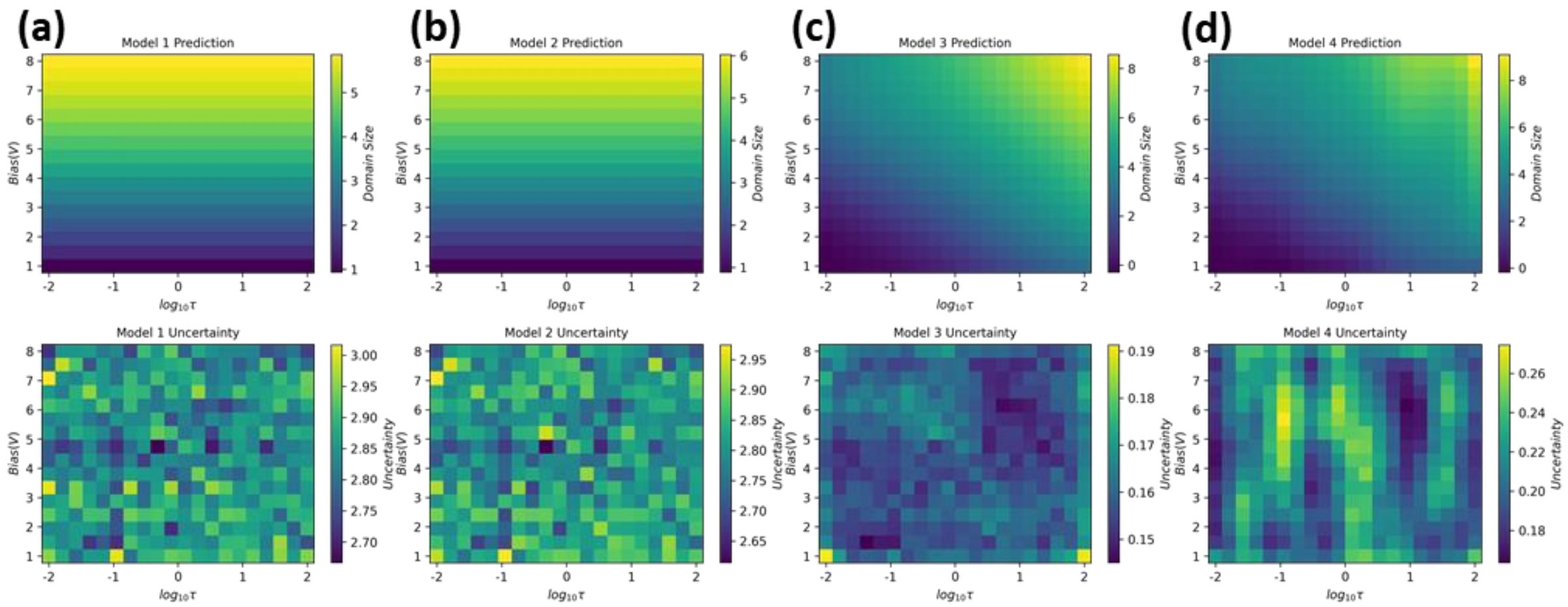

**Figure 6.** Predictions by all models on the final set of discovered parameters after the competition of the experiment. (**a-d**) Prediction (top row) and corresponding uncertainties (bottom row) by four different models, respectively.

**Table 2.** Final parameters of each model obtained from hypotheses-learning automated experiment.

<table>
<tr><th></th><th>Model Equation</th><th colspan="18">Final Model Parameters</th></tr>
<tr><td rowspan="3"><b>Model I</b></td><td rowspan="3">$r(V) = r_{cr} + r_0\sqrt{\left(\frac{V}{V_c}\right)^{2/3} - 1}$</td><td colspan="6">$r_{cr}$</td><td colspan="6">$r_0$</td><td colspan="6">$V_c$</td></tr>
<tr><td colspan="2">$\bar{x}$</td><td colspan="2">M</td><td colspan="2">s</td><td colspan="2">$\bar{x}$</td><td colspan="2">M</td><td colspan="2">s</td><td colspan="2">$\bar{x}$</td><td colspan="2">M</td><td colspan="2">s</td></tr>
<tr><td colspan="2">-0.23</td><td colspan="2">-0.25</td><td colspan="2">0.89</td><td colspan="2">2.35</td><td colspan="2">2.37</td><td colspan="2">0.72</td><td colspan="2">0.53</td><td colspan="2">0.53</td><td colspan="2">0.24</td></tr>
<tr><td rowspan="3"><b>Model II</b></td><td rowspan="3">$r(V) = r_{cr} + r_0\sqrt[3]{\left(\frac{V}{V_c}\right)^{2} - 1}$</td><td colspan="6">$r_{cr}$</td><td colspan="6">$r_0$</td><td colspan="6">$V_c$</td></tr>
<tr><td colspan="2">$\bar{x}$</td><td colspan="2">M</td><td colspan="2">s</td><td colspan="2">$\bar{x}$</td><td colspan="2">M</td><td colspan="2">s</td><td colspan="2">$\bar{x}$</td><td colspan="2">M</td><td colspan="2">s</td></tr>
<tr><td colspan="2">0.06</td><td colspan="2">0.84</td><td colspan="2">0.02</td><td colspan="2">1.03</td><td colspan="2">1.03</td><td colspan="2">0.36</td><td colspan="2">0.65</td><td colspan="2">0.59</td><td colspan="2">0.40</td></tr>
<tr><td rowspan="3"><b>Model III</b></td><td rowspan="3">$r(V,t) = V^{\alpha}\log\tau$</td><td colspan="18">$\alpha$</td></tr>
<tr><td colspan="6">$\bar{x}$</td><td colspan="6">M</td><td colspan="6">s</td></tr>
<tr><td colspan="6">0.38</td><td colspan="6">0.37</td><td colspan="6">0.05</td></tr>
<tr><td rowspan="2"><b>Model IV</b></td><td rowspan="2">$r(V,t) = V^{\alpha}\tau^{\beta}$</td><td colspan="9">$\alpha$</td><td colspan="9">$\beta$</td></tr>
<tr><td colspan="3">$\bar{x}$</td><td colspan="3">M</td><td colspan="3">s</td><td colspan="3">$\bar{x}$</td><td colspan="3">M</td><td colspan="3">s</td></tr>
</table>

| | | | | | | | |
|---|---|---|---|---|---|---|---|
| | | 0.81 | 0.81 | 0.01 | 0.33 | 0.33 | 0.00 |

*Table Notes:*

*(1)* $\bar{x}$*: mean value;* M*: median value;* s*: standard deviation.*

*(2) All model parameters are obtained from the last time a particular model was sampled in the hypotheses-learning experiment. Model I parameters are obtained from training step 38, measurement step 58; Model II parameters are obtained from training step 22, measurement step 42; Model III parameters are obtained from training step 40, measurement step 60; Model IV parameters are obtained from training step 39, measurement step 59.*

Shown in Figure 6 are predictions and uncertainties by four models after the experiment based on all obtained results. The predictions of models 3 and 4 describe the experimental results (Figure 5b) better, and model 3 exhibits the lowest uncertainty. These results suggest that the domain growth in this BTO thin film is determined by the kinetics of the domain wall motion rather than thermodynamics of the screening process, and the effect of surface screen charges is minor. In the automated experiment, the hypothesis learning also actively updates the model parameters. At the competition of the experiment, each model got parameters that best describe the experimental data. The final parameters of each model are summarized in Table 2.

To summarize, we have illustrated the hypothesis learning based automated experiment to explore domain switching. The hypothesis list has been built for different domain growth limiting stages, including thermodynamics of domain formation, domain wall pinning, and transport of screening charges at the surfaces. The results indicate that domain growth is ruled by kinetic control.

This approach for probing local bias-induced transformation is general and can be used for other tip-induced reactions and processes, including reversible and irreversible tip-induced electrochemical reactions including electroplating[75] and nano oxidation.[76, 77] Note that the detection signal is not limited to the direct measurement of the domain size, and can include measured currents, changes in topography, and resonance frequency shifts. As such, it can provide a powerful tool for probing neuromorphic materials,[78] fuel cell and battery materials,[79] as well as provide fundamental insights into electrochemical processes on the nanometer scale.[80]

We further note that the hypothesis learning can be broadly used in other automated experiment settings. Currently, this includes exploration of the relatively low-dimensional parameter cases for which easy to evaluate competing physical models are available such as automated synthesis via microfluidic and robotic systems,[81] pulsed laser deposition, and other forms of materials synthesis, etc.

**Contributions:** YL implemented the hypothesis learning microscopy and collected and analyzed the data. MZ realized the hypothesis learning algorithms. KPK developed the LabView script for FPGA. SVK proposed and led the research. AM developed the analytical models. The authors acknowledge Stephen Jesse for helpful discussions and implementation of hardware-software integration and Vladimir A. Protopopescu for helpful discussions. All authors participated in the discussion.

**Acknowledgements:** This effort (Y.L., S.V.K.) was supported as part of the center for 3D Ferroelectric Microelectronics (3DFeM), an Energy Frontier Research Center funded by the U.S. Department of Energy (DOE), Office of Science, Basic Energy Sciences under Award Number DE-SC0021118. The research was performed and partially supported (M.Z.) at Oak Ridge National Laboratory's Center for Nanophase Materials Sciences (CNMS), a U.S. Department of Energy, Office of Science User Facility. A.N.M. was supported by the National Academy of Sciences of Ukraine (the Target Program of Basic Research of the National Academy of Sciences of Ukraine "Prospective basic research and innovative development of nanomaterials and nanotechnologies for 2020 - 2024", Project № 1/20-H, state registration number: 0120U102306) and received funding from the European Union's Horizon 2020 research and innovation programme under the Marie Skłodowska-Curie grant agreement No 778070.

## Materials and methods:

*Materials:*

$BaTiO_3$ thin films were grown via pulsed laser deposition (PLD) in 99.9999% pure $O_2$ at 700˚C. Specifically, first a 5 nm $SrRuO_3$ back electrode was grown on (100) single sided epitaxial-polished $SrTiO_3$ substrates at 100 millitorr with a pulse rate of 5 Hz from a stoichiometric $SrRuO_3$ ceramic target. Subsequently, 80 nm of $BaTiO_3$ was grown at 10 millitorr with a laser pulse rate of 10 Hz from a stoichiometric BaTiO3 target. The fluence for both thin film layers was maintained at approximately 1.2 J/cm$^2$. Substrates were prepared by sonication in a warm (~70 °C) deionized water bath for 1 min followed by an anneal at 1000 °C for 12 hours to produce $TiO_2$ termination with step and terrace surface morphology.

*Automated experiment in PFM:*

The hypothesis learning driven automated BEPFM measurement is based on an Asylum Research Cypher microscope equipped with a National Instruments DAQ card with LabView and a Field Programmable Gate Arrays with Python Jupyter Notebook. For domain writing, FPGA moves the tip to a desired location and apply a DC bias. Followed by a BEPFM image measurement performed with NI, FPGA, and Cypher. the FPGA performs scan (move tip) and send to trigger to NI DAQ card to perform BE measurement simultaneously. These processes are embedded in a Jupyter Notebook. When a measurement finishes, the Jupyter Notebook analyzes the BEPFM phase image to obtain the domain size and saves domain size in Google Drive. Then, the hypothesis training is performed in Google Colaboratory with this domain size (and previous domain size), followed by saving the next writing parameters for next experiment.

*Hypothesis learning:*

The hypothesis learning (hypoAL)[35] was implemented using the home-build GPax package:https://github.com/ziatdinovmax/gpax. The probabilistic models were wrapped into the structured Gaussian processes[36] and the Bayesian inference was performed via the iterative No-U-Turn sampler[82]. To ensure that wrapped models 1 and 2 remained isotropic in time, the kernel lengthscale for the time dimension was set to a sufficiently large value (1000), whereas the kernel lengthscale for the voltage dimension was sampled from a standard weakly informative log-normal prior. For models 3 and 4, the ARD kernel in both dimensions was sampled from log-normal priors.

The acquisition function value in each unmeasured point $x_*$ was equal to the posterior predictive uncertainty

$$\mathbb{V}[f_*] = \frac{1}{N}\sum_{n=1}^{N}(f_*^n - \hat{f}_*)^2, \text{ with } \hat{f}_* = \frac{1}{N}\sum_{n=1}^{N} P(x_*|\theta^n, D)$$

where $\theta^n \sim P(\theta|D)$ were samples drawn from the posterior and $D$ was the available (measured) data. The reward function was defined as

$$R\left(\mathbb{V}_{\mathrm{m}}^{i}, \mathbb{V}_{\mathrm{m}}^{i-1}\right) = \begin{cases} +1, & \mathbb{V}_{\mathrm{m}}^{i} < \mathbb{V}_{\mathrm{m}}^{i-1} \\ -1, & \mathbb{V}_{\mathrm{m}}^{i} \geq \mathbb{V}_{\mathrm{m}}^{i-1} \end{cases},$$

where $\mathbb{V}_{\mathrm{m}}^{i}$ is a median value of posterior predictive uncertainty at step $i$. The python script used to run the hypoAL during the experiment can be found in the Supplementary Material.